\documentclass[prl,aps,epsf,floats,showpacs]{revtex4}

\usepackage{graphicx}

\begin{document}

\title{ Two-dimensional quantum dot helium in a magnetic field: 
        variational theory }

\author{Orion Ciftja}

\affiliation{
         Department of Physics,
         Prairie View A\&M University, Prairie View, Texas 77446 }

\author{M. Golam Faruk}

\affiliation{
         Department of Electrical Engineering and
         Department of Physics,
         Prairie View A\&M University, Prairie View, Texas 77446 }


\begin{abstract}

A trial wave function for two-dimensional quantum dot helium in an 
arbitrary perpendicular magnetic field (a system of two interacting electrons 
in a two-dimensional parabolic confinement potential) is introduced.
A key ingredient of this trial wave function is a Jastrow pair  correlation 
factor that has a displaced Gaussian form.
The above choice of the pair correlation factor is instrumental on assuring 
the overall quality of the wave function at all values of the magnetic field.
Exact numerical diagonalization results are used to gauge the
quality of the proposed trial wave function.
We find out that this trial wave function is an excellent representation of 
the true ground state at all values of the magnetic field including 
weak (or zero) and strong magnetic fields.

\end{abstract}



\maketitle

\section{ Introduction }
\label{sec1}

Among the variety of two-dimensional (2D) few electron quantum 
dots~\cite{Jacak97,Bryant87,Ashoori92,Ashoori96,Kouwenhoven97,Heitmann93,Kastner,Tarucha,Tar96,Sasaki,Aus99,Peeters96,pet1,pet2},
the two-electron ($N=2$) quantum dot stands out as a truly remarkable system that, despite
the simplicity, shows very rich phenomena and posseses characteristic 
features which persist to larger systems.
Such a system 
oftenly refered to as  2D {\it quantum dot helium} exhibits
a highly complex behavior in presence of a perpendicular magnetic field.
Its ground state energetics is unusually complicated and intricate 
singlet to triplet spin state transitions
occur as the magnetic field is varied.~\cite{wagner,singlettriplet}

The Hamiltonian of 2D quantum dot helium in a perpendicular magnetic field 
can be written as:

\begin{eqnarray} 
\hat{H}(\vec{\rho}_{1},\vec{\rho}_{2}) &=&
\sum_{i=1}^{2} \left \{ \frac{p_i^2}{2 m}+\frac{\omega_c}{2} \hat{L}_{iz}+
\frac{m}{2} \, \left[ \omega_0^2+\left(\frac{\omega_c}{2}\right)^2 \right] \, 
\rho_i^2 \right \} +
\nonumber \\ & &
+\frac{1}{4 \pi \epsilon_0 \epsilon_r} 
\frac{e^2}{|\vec{\rho}_1-\vec{\rho}_2|}+
g_e \, \mu_B \, B_z \, S_z \ ,  
\label{twoosci}
\end{eqnarray} 
where
$\hat{\vec{p}}_i=(\hat{p}_{ix},\hat{p}_{iy})$ and
$\vec{\rho}_i=(\rho_{ix},\rho_{iy})$ are, respectively,
the 2D momentum and position of the $i$-th electron,
$m$ is electron's mass,
$-e(e>0)$ is electron's charge, 
$g_e$ is electron's g-factor, 
$\mu_B$ is Bohr's magneton,
$\epsilon_r$ is the dielectric constant,
$\hbar \, \omega_0$ is the parabolic confinement energy, 
$S_z$ is the $z$-component of the total spin,
$B_z$ is the perpendicular magnetic field,
$\hat{L}_{iz}$ is the $z$-component angular momentum operator for
the $i$-th electron and
 $\omega_c=e B_z/m>0$ is the cyclotron frequency.

%

There have been several studies of 2D quantum dot helium with or without 
magnetic field employing a variety of techniques.
For example,
Merkt et al.~\cite{Merkt91} studied the energy spectra of two interacting
electrons in a parabolic potential in absence and in presence of
a perpendicular magnetic field employing the 
exact numerical diagonalization technique.
Wagner et al.~\cite{wagner} used the exact numerical 
diagonalization technique to study 2D quantum dot helium in a perpendicular
magnetic field and to
predict oscillations between spin-singlet and spin-triplet 
states as a function of the magnetic field strength.
Pfannkuche et al.~\cite{Pfannkuche93} compared energies, pair correlation
functions, and particle densities 
obtained from the Hartree, Hartree-Fock (HF), and 
exact numerical diagonalization method and
pointed out the unsuitability of the HF approach at weak
magnetic fields.
Pfannkuche et al.~\cite{Pfannkuche_physicab} 
used the diagonalization data to formulate a theory of
2D quantum dot helium in a perpendicular magnetic field, explaining how
the competition between the Coulomb interaction and the binding
forces due to confinement and the magnetic field induces 
ground state transitions.
Harju et al.~\cite{HarjuphysicaB} 
introduced a recipe of how to build a trial wave function
for 2D quantum dot helium in a perpendicular magnetic field
and applied the variational Monte Carlo (VMC) technique.
His Jastrow pair correlation factor has two variational parameters
and takes into account the mixing of different Landau levels for 
the relative motion, though the ansatz does not have the right asymptotic 
behavior at large interparticle distances.

A wealth of information in 2D quantum dots, in general, and 2D quantum 
dot helium in particular, is provided by the rotating Wigner (or electron) 
molecule (RWM or REM) theory.~\cite{maksymtheory,uzi1}
Specific REM trial wave functions have been recently derived~\cite{uzi2} for
2D quantum dot systems in high magnetic fields.
The REM wave functions are constructed by first breaking the rotational 
symmetry at the unrestricted Hartree-Fock (UHF) level and, secondly, 
restoring the circular symmetry via post HF methods and projection 
techniques.~\cite{uzi1}
The UHF-level REM wave function describes a Wigner molecule that is considered
as a rigid rotor, while the second step of restoring the circular 
symmetry implies rotations of such molecules.
At zero and weak magnetic fields the broken-symmetry UHF orbitals need be 
determined numerically while at high magnetic fields they can be well 
approximated by one-particle lowest Landau level (LLL) Gaussian functions 
that are centered at positions that correspond
to the classical equilibrium configuration of $N$ point charges in a 
harmonic trap.
Yannouleas and Landman~\cite{uzi1} used such RWM trial wave function to 
study 2D quantum dot helium in a magnetic field. 
They calculated the ground state energies as a function
of the magnetic field for the spin triplet states of 2D quantum dot helium 
and noted that for high magnetic fields ( larger that 7 T, see Fig. 1 
in Ref.~\cite{uzi1}) the RWM and exact diagonalization results are 
practically the same.

Some of the main theoretical methods used to study quantum
dots are:
analytical calculations \cite{Taut93,Taut94,Turbiner94,Dineykhan97}, 
exact numerical 
diagonalizations~\cite{Maksym90,MacDonald93,Yannouleas00,Eto97},
quantum Monte Carlo (QMC) methods \cite{Maksym96,Bolton96,Kainz02,Harju02},
density functional 
theory methods~\cite{Kos97,Hir99,Steffens98a,Steffens98b,Fer94}
and
Hartree-Fock mean-field 
theory.~\cite{Kum90,Fuj96,Mul96,Yan99}
These methods have advantages and disadvantages, most notably some methods
that deliver good results for the ground state properties may be
inadequate in describing excited state properties and vice versa.
Here we are limiting our discussion to the calculation of ground state
properties only.
Considering the limitations of several methods on obtaining ground state
properties (for example: exact numerical diagonalization methods are 
applicable only for 
few electrons,  Hartree-Fock and perturbation theory lack the
desired accuracy, etc), the use of QMC methods,
such as the VMC technique seems to be the best strategy in the long run.
Therefore the quest for better, yet simple, trial ground state wave functions 
is always timely.
A high quality trial ground state wave function is essential not only for the
VMC method, but also for 
the more sophisticated diffusion Monte Carlo (DMC) method that
relies on a guiding trial wave function.
Compared to other methods, QMC methods have the greatest advantage of all, 
since they can be extended to larger number of electrons in a 
straightforward manner and are very accurate.

In this work we introduce a novel trial wave function to describe 
the ground state of 2D quantum dot helium in a perpendicular magnetic field. 
Such a wave function is written as a product of a Laughlin-type 
wave function~\cite{laughlin,eswf,efhnc}
with a Jastrow pair correlation factor and has the form:
\begin{equation}
\Psi(\vec{\rho}_1,\vec{\rho}_2)=J(\rho_{12}) \times (z_1-z_2)^{|m_z|} \, 
\exp{\left( -\frac{\rho_1^2+\rho_2^2}{4 \, l_{\Omega}^2} \right)}  \ ,
\label{jastrow_laughlin}
\end{equation}
where the Jastrow factor, $J(\rho_{12})$ is:
\begin{equation}
J(\rho_{12})=\exp{\left( -\frac{b^2}{2} \, \rho_{12}^2+
               c \, b \, \rho_{12}  \right)} \ .
\label{jasmanybodywf}
\end{equation}
The 2D position coordinate, $z_j=x_j-i y_j$ of the Laughlin component of 
the wave function is given in complex notation, 
$\rho_{12}=|\vec{\rho}_1-\vec{\rho}_2|$ is the inter-electron distance,
$m_z=|m_z|=0,1,\ldots$ is the angular momentum number,
$b$ and $c$ are non-negative variational parameters to be optimized.
We have
$l_{\Omega}=\sqrt{\hbar/(2 m \Omega)}$ and 
$\Omega^2=\omega_0^2+(\omega_c/2)^2$.
The effective magnetic length, $l_{\Omega}$ reduces to the electronic
magnetic length, $l_0=\sqrt{\frac{\hbar}{e \, B_z}}$ when there is 
no confinement ($\omega_0=0$) or when the magnetic field is very
large ( $\omega_c/\omega_0 \rightarrow \infty$ ).
The effective magnetic length, $l_{\Omega}$ can be written in terms
of the inverse oscillator length, $\alpha=\sqrt{m \omega_0/\hbar}$ as:
\begin{equation} 
\frac{1}{l_{\Omega}^2}=2 \, \alpha^2 \, \sqrt{1+\frac{1}{4} 
\left(\frac{\omega_c}{\omega_0}\right)^2} \ .
\label{lomega}
\end{equation}
In this way it is easy to recover the 2D harmonic oscillator states
in the limit of zero magnetic field 
( $\omega_c/\omega_0 \rightarrow 0$).
The parity of the ground state space wave function depends on the value of
angular momentum $|m_z|$.
For even values, $|m_z|=0,2,4,\ldots$ the space wave function is symmetric
and the spin function corresponds to a spin-singlet
state ($S=0$), while for odd values, $|m_z|=1,3,5,\ldots$ the 
space wave function is antisymmetric and the spin function
becomes a spin-triplet ($S=1$).

What makes this trial wave function rather unique is its
displaced Gaussian Jastrow pair correlation factor,
$J(\rho_{12})$,
which is different from earlier choices in the 
literature.~\cite{Bolton96,HarjuphysicaB,pederiva}
The rationale behind the choice displaced Gaussian choice of correlation
factor can be better understood if one considers a pair 
of electrons in zero magnetic field.
In absence of an electronic repulsion between the electrons
the relative coordinate ground state wave function 
with be a Gaussian centered at coordinate, 
$\rho_{12}=0$, will have zero angular momentum and
will correspond to a spin singlet state.
With the Coulomb repulsion
the ground state will still have zero angular momentum~\cite{Bolton96}
therefore it is plausible to expect that the
main effect of the Coulomb correlation is simply to further separate
the electrons resulting in a new relative coordinate ground state wave function
which will resemble a Gaussian centered at $\rho_{12} \neq 0$ values.
The choice in Eq.(\ref{jasmanybodywf}) mimicks this physical effect.

Trial wave functions, such as the REM wave functions (for $N=2$ and $N>2$ 
electrons) also take into consideration the relative separation of electrons 
and have led to a rich physics regarding Wigner crystallization and the 
rotation of the electron molecules formed in high magnetic 
fields.~\cite{uzi1,uzi3}
Within the framework of the REM wave functions, which by construction are 
crystalline in character, the separation between electrons is achieved at 
the one-particle level right at the start.
For instance, in the high magnetic field regime, the REM wave function's 
Slater determinants contains LLL one-particle Gaussian orbitals 
which are centered at different positions, $Z_j$ (in complex notation) and 
have the form:
\begin{equation}
u(z,Z_j)=\frac{1}{2 \, \pi \, l_0^2} \, 
          \exp\left[ -\frac{|z-Z_j|^2}{4 \, l_0^2} 
           -\frac{i}{2 \, l_0^2} \, (x \, Y_j-y \, X_j) \right] \,
\label{lllgaussian}
\end{equation} 
where $Z_j$ coincide with the classical equilibrium positions of classical
point charges and $l_0$ is the electronic magnetic length.
In the case of our variational wave function, the separation between electrons
is achieved at the two-particle (pair) level through the displaced Gaussian 
pair correlation factor.
Despite the same idea of optimizing the separation between electrons, the 
displaced Gaussian pair correlation factors considered in this work have 
a two-body
structure which makes it different from the one-particle (displaced) Gaussian
functions that approximate the one-particle UHF orbitals of the REM wave 
functions at high magnetic fields.~\cite{uzi2}

Despite the usual controversies involved on the 
selection of a Jastrow correlation factor
it appears clearly that the displaced Gaussian 
pair correlation factor has all the attributes to capture 
effectively most of the electronic correlations present
on the ground state of such system.
Some indication of the eventual high quality 
of the trial wave function also comes from a previous study of 2D 
quantum dot helium at zero magnetic field~\cite{oriondot} where
the same Jastrow pair correlation factor was used.

Motivated by these arguments, it is the objective of this work
to perform a complete study of 2D quantum dot helium 
in an arbitrary perpendicular magnetic field that ranges from very
weak (or zero) to very strong
using the trial wave function introduced in Eq.(\ref{jastrow_laughlin})
and Eq.(\ref{jasmanybodywf}).
In the process we will test how effective the displaced Gaussian 
pair correlation factor is on representing the effect of 
electronic correlations at arbitrary magnetic fields.
Exact numerical diagonalization calculations will be used as a
gauge of accuracy of the variational results.

The paper is organized as follows:
In Section~\ref{sec_var} we introduce the variational method,
the trial wave function and calculate various quantities corresponding
to such a wave function.
In Section~\ref{results} we show the results obtained from the variational
wave function and compare them to the exact diagonalization method.
A discussion of the results is given in Section~\ref{discussion} 
and concluding remarks can be found in Section~\ref{summary}.

\section{ Variational theory }
\label{sec_var}

In this section we apply the variational method to study 
2D quantum dot helium in a perpendicular magnetic field using
the trial wave function of Eq.(\ref{jastrow_laughlin}) with the 
Jastrow pair correlation factor having the displaced
Gaussian form given in Eq.(\ref{jasmanybodywf}).
We will show that, after optimization, the
proposed trial wave function is an excellent representation of the
true ground state at any value of the magnetic field and compares very 
favorably to the exact numerical diagonalization results.

There are two dimensionless parameters that determine the behavior
and properties of the system under consideration:
\begin{equation}
\lambda=\frac{1}{4 \pi \epsilon_0 \epsilon_r}
\frac{e^2 \, \alpha}{ \hbar \omega_0} \ \ \ ; \ \ \ 
       \gamma=\frac{\omega_c}{\omega_0} \ ,
\label{definelam}
\end{equation}
where $\lambda$ gauges the strength of the Coulomb correlation relative
to the confinement energy and $\gamma$ measures the strength of the
magnetic field relative to confinement.
One cam immediately see that 
$\lambda=l/a_B$, where $l=1/\alpha$ is the harmonic oscillator length
and
$a_B=4 \pi \epsilon_0 \epsilon_r \hbar^2/(m e^2)$ is the effective
Bohr radius.

Since the parity of space wave function is determined
by the value of the angular momentum, $|m_z|$, the ground state
angular momentum value determines whether the ground state
corresponds to a spin-singlet or spin-triplet state.
Therefore, a stringent test of quality for this trial wave 
function is to check whether the lowest variational energy state has
always the correct angular momentum number 
as calculated from the numerical diagonalizations
under different combinations of Coulomb correlation, 
confinement and magnetic field. 

In a general situation where both Coulomb correlation and 
magnetic field are present, there is no way to anticipate the 
correct value of the ground state angular momentum number.
An exception are the simple cases of:
(i) absence of Coulomb correlations 
or 
(ii) absence of a perpendicular magnetic field,
where it is straightforward to prove that the ground state is expected 
to have zero angular momentum.

Obviously, such behavior should be reflected by the trial wave function
under investigation.
With no Coulomb correlation ($\lambda=0$), the ground state has
zero angular momemtum ($|m_z|=0$) both in presence
or absence of the perpendicular magnetic field.
Under these conditions, the corresponding trial wave function 
becomes that of Eq.(~\ref{jastrow_laughlin}) with 
Jastrow correlation $J(\rho_{12})=1$ and angular momentum, $|m_z|=0$
as expected.
In absence of a perpendicular magnetic field, the groundstate still has zero
angular momentum, ($|m_z|=0$) with or without the Coulomb correlation,
therefore a trial wave function with 
$J(\rho_{12}) \neq 1$ and angular momentum, $|m_z|=0$ is again
consistent with the expected scenario.

However,
when both Coulomb correlation and magnetic field are present, the
situation changes drastically.
As the magnetic field increases, a groundstate with nonzero
angular momentums ($|m_z| \ne 0$) may arise, therefore in addition 
to the Jastrow pair correlation factor also the 
Laughlin factor starts contributing on keeping the electrons apart 
in a more effective way.

A calculation of the expectation value of the Hamiltonian with 
respect to the trial wave function,
$E=\langle \Psi |\hat{H}|\Psi \rangle/\langle \Psi |\Psi \rangle$
gives:

\begin{eqnarray} 
\epsilon(B,c,\gamma,m_z,\lambda) &=&
=\frac{E}{\hbar \omega_0}=
-\frac{|m_z|}{2} \, \gamma+
  \nonumber \\ & &
\frac{-B^2 \, f_1(B,c,\gamma,m_z)+(1+\frac{\gamma^2}{4}) \, 
\frac{f_2(B,c,\gamma,m_z)}{(4 B^2)}+
         \lambda \, B \, f_3(B,c,\gamma,m_z)}{f(B,c,\gamma,m_z)}
  \nonumber \\ & &
+\sqrt{1+\frac{\gamma^2}{4}}    \ ,
\label{eexpres}
\end{eqnarray}
where
$B=b/\alpha$ and $c$ are variational parameters,
$\gamma=\omega_c/\omega_0$ 
is linearly proportional to the magnetic field $(\propto B_z)$ 
and 
$\lambda=e^2 \alpha/(4 \pi \epsilon_0 \epsilon_r \hbar \omega_0)$
is the Coulomb correlation parameter.
All the above quantities are dimensionless and the ground state energy 
is given in units of $\hbar \, \omega_0$.
Note that the Zeeman energy term is not specifically included in the
expression for the variational energy.

The functions: $f_{1,2,3}$ and $f$, as well as the function $g$, depend on the 
variables specified on their arguments and in integral form are given by:

\begin{equation}
\left\{
\begin{array}{l}
f_1(B,c,\gamma,m_z)=\int_{0}^{\infty}dt \, t \, g(t,B,c,\gamma,m_z) \times \\
\times    
\{ \left[\frac{|m_z|}{t}+c-t \left(1+\frac{1}{2 B^2} 
          \sqrt{1+\frac{\gamma^2}{4}} \right) \right]^2 -\frac{1}{B^2}
          \sqrt{1+\frac{\gamma^2}{4}}-2  \\ 
         +\frac{c}{t}-\frac{|m_z|^2}{t^2} \} \ , \\  \\
f_2(B,c,\gamma,m_z)=\int_{0}^{\infty}dt \, t^3 \, g(t,B,c,\gamma,m_z)
               \ , \\ \\
f_3(B,c,\gamma,m_z)=\int_{0}^{\infty}dt \, g(t,B,c,\gamma,m_z)
                \ , \\ \\ 
f(B,c,\gamma,m_z)=\int_{0}^{\infty}dt \, t \, g(t,B,c,\gamma,m_z)
             \ , \\ \\ 
g(t,B,c,\gamma,m_z)=t^{2 |m_z|} \, 
e^{-\frac{t^2}{2 B^2} \sqrt{1+\frac{\gamma^2}{4}}-t^2+2 \, c \, t}   \ ,
\label{fourfun}
\end{array}
\right.
\end{equation}
where $t=b \, \rho_{12}$ 
is an auxiliary variable introduced to simplify the calculation of integrals.
The optimization procedure is straightforward:
given the values of Coulomb and magnetic field parameters, 
$\lambda$ and $\gamma$ 
we calculate the lowest energies for a set of integer values of $m_z$
by optimizing the variational parameters, $B$, and $c$ through
standard numerical procedures.

\section{ Results}

\label{results}

The best way to gauge the accuracy of the trial wave function
is to directly compare the variational results to exact numerical
diagonalization values.
%
%
Table~\ref{tabdiag} displays the diagonalization
ground state energies, $\epsilon=E/(\hbar \omega_0)$ for 2D quantum dot 
helium in a perpendicular magnetic field for values of Coulomb
correlation,
     $\lambda=1, \ldots 6$,
and values of magnetic field,
     $\gamma=0,1,\ldots, 5$.
The ground state angular momentum, $m_z$ is also specified.

\begin{table}[!t]
\caption[ ]{ The exact numerical diagonalization 
            ground state energies, $\epsilon=E/(\hbar \omega_0)$ 
            for 2D quantum dot helium subject to a perpendicular
            magnetic field as a function of dimensionless Coulomb 
            coupling parameter, 
            $\lambda=e^2 \alpha/(4 \pi \epsilon_0 \epsilon_r \hbar \omega_0)=0,
              1, \ldots 6$
            and values of magnetic field,
            $\gamma=\omega_c/\omega_0=0,1,\ldots, 5$.
            The angular momentum, $m_z$ of the ground state is also
            specified.
            The parameter
            $ \alpha=\sqrt{m \, \omega_0/\hbar}$ has the dimensionality
            of an inverse length.   }
\label{tabdiag}
\begin{center}
\begin{tabular}{|l|l|l|l|l|l|l|}
\hline                                                  

 & $\gamma=0$ & $\gamma=1$ & $\gamma=2$ & $\gamma=3$ 
           & $\gamma=4$ & $\gamma=5$                      \\ \hline
$\lambda=0$ & 2.00000 & 2.23607 & 2.82843  & 3.60555  & 4.47214   & 5.38516 
                                                            \\ 
  &$m_z=0$ & $m_z=0$ & $m_z=0$ & $m_z=0$ & $m_z=0$ & $m_z=0$  \\ \hline
$\lambda=1$  & 3.00097 & 3.30508 & 3.95732  & 4.71894  & 5.61430  & 6.53067 
                                                            \\ 
 &$m_z=0$ & $m_z=0$ & $m_z=1$ & $m_z=1$ & $m_z=1$ & $m_z=2$   \\ \hline
$\lambda=2$ & 3.72143 & 4.06684 & 4.61879  & 5.43123  & 6.30766  & 7.22681
                                                            \\ 
 & $m_z=0$ & $m_z=1$ & $m_z=1$ & $m_z=2$ & $m_z=2$ & $m_z=3$ \\ \hline
$\lambda=3$ & 4.31872 & 4.60594 & 5.23689  & 6.01256  & 6.89002  & 7.81384 
                                                            \\ 
 & $m_z=0$ & $m_z=1$ & $m_z=1$ & $m_z=2$ & $m_z=3$ & $m_z=4$  \\ \hline
$\lambda=4$ & 4.84780 & 5.11165 & 5.73642  & 6.53522  & 7.41600  & 8.33874 
                                                            \\ 
  & $m_z=0$ & $m_z=1$ & $m_z=2$ & $m_z=3$ & $m_z=4$ & $m_z=5$   \\ \hline
$\lambda=5$ & 5.33224 & 5.58995 & 6.21499  & 7.01716  & 7.90109  & 8.82281 
                                                            \\ 
 & $m_z=0$ & $m_z=1$ & $m_z=2$ & $m_z=3$ & $m_z=4$ & $m_z=6$  \\ \hline
$\lambda=6$ & 5.78429 & 6.04534 & 6.67999  & 7.46782  & 8.34530  & 9.27057 
                                                            \\ 
 & $m_z=0$ & $m_z=1$ & $m_z=2$ & $m_z=4$ & $m_z=5$ & $m_z=6$  \\ \hline
\end{tabular}
\end{center}
\end{table}

%
%
In Table~\ref{tabvar} we show the variational ground state energies
and optimal values of parameters, $B$ and $c$
for 2D quantum dot helium in a perpendicular magnetic field 
at different  $\lambda$-s and $\gamma$-s.
The 
results are rounded in the last digit.
%
%
\begin{table}[!t]
\caption[]{ The variational
            ground state energies, $\epsilon=E/(\hbar \omega_0)$,
            angular momentum values, $m_z$, as well as optimal
            parameter values, $B=b/\alpha$ and $c$ 
            for 2D quantum dot helium subject to a perpendicular
            magnetic field
            as a function of dimensionless Coulomb coupling parameter, 
            $\lambda=e^2 \alpha/(4 \pi \epsilon_0 \epsilon_r \hbar \omega_0)=0,
              1, \ldots 6 $ 
            and values of magnetic field,
            $\gamma=\omega_c/\omega_0=0,1,\ldots, 5$.
            The parameter
            $ \alpha=\sqrt{m \, \omega_0/\hbar}$ has the dimensionality
            of an inverse length.   }
\label{tabvar}
\begin{center}
\begin{tabular}{|l|l|l|l|l|l|l|}
\hline                                                  
& $\gamma=0$ & $\gamma=1$ & $\gamma=2$ & $\gamma=3$ 
           & $\gamma=4$ & $\gamma=5$ 
                                                        \\ \hline
$\lambda=0$ & 2.00000 & 2.23607 & 2.82843  & 3.60555  & 4.47214   & 5.38516 
                                                            \\ 
$m_z$       & 0       & 0       & 0        & 0        & 0        & 0    \\ 
$B$         & 0       & 0       & 0        & 0        & 0        & 0    \\ 
$c$         & 0       & 0       & 0        & 0        & 0        & 0    \\  \hline
$\lambda=1$ & 3.00174 & 3.30578 & 3.95737  & 4.71899  & 5.61435  & 6.53068  
                                                            \\ 

$m_z$       & 0        & 0        & 1        & 1        & 1        & 2          \\ 
$B$         & 0.40185  & 0.41627  & 0.22809  & 0.24247  & 0.25600  & 0.18816   \\ 
$c$         & 1.67676  & 1.63743  & 1.11252  & 1.05000  & 0.99691  & 0.81641   \\ \hline

$\lambda=2$ & 3.72565 & 4.06704 & 4.61899  & 5.43127  & 6.30769  & 7.22683   
                                                            \\ 
$m_z$       & 0        & 1        & 1        & 2        & 2        & 3           \\ 
$B$         & 0.49791  & 0.29039  & 0.30813  & 0.24041  & 0.25184  & 0.20584    \\     
$c$         & 2.21655  & 1.64297  & 1.55874  & 1.26233  & 1.20167  & 1.04900     \\ \hline     

$\lambda=3$ & 4.32576 & 4.60635 & 5.23732  & 6.01263  & 6.89004  & 7.81385   
                                                            \\ 
$m_z$       & 0        & 1        & 1        & 2        & 3        & 4          \\ 
$B$         & 0.54243  & 0.34137  & 0.36669  & 0.28354  & 0.23794  & 0.21288    \\ 
$c$         & 2.57492  & 1.98559  & 1.88097  & 1.54922  & 1.34215  & 1.18987     \\ \hline

$\lambda=4$ & 4.85637  & 5.11233  & 5.73655  & 6.53525  & 7.41601  & 8.33875   
                                                            \\ 
$m_z$       & 0        & 1        & 2        & 3        & 4        & 5           \\  
$B$         & 0.56603  & 0.37934  & 0.30103  & 0.25770  & 0.22591  & 0.20806    \\ 
$c$         & 2.85052  & 2.26315  & 1.88523  & 1.62705  & 1.45132  & 1.30691     \\ \hline

$\lambda=5$ & 5.34141  & 5.59088 & 6.21518  & 7.01722  & 7.90111  & 8.82282   
                                                            \\ 
$m_z$       & 0        & 1        & 2        & 3        & 4         & 6          \\ 
$B$         & 0.57976  & 0.40862  & 0.33005  & 0.28447  & 0.25575   & 0.20855   \\ 
$c$         & 3.07820  & 2.49976  & 2.09758  & 1.81530  & 1.60855   & 1.38742    \\ \hline

$\lambda=6$ & 5.79354  & 6.04652 & 6.68025  & 7.46785  & 8.34532  & 9.27058   
                                                            \\ 
$m_z$       & 0        & 1        & 2        & 4        & 5         & 6           \\  
$B$         & 0.58818  & 0.43192  & 0.35512  & 0.26257  & 0.24321   & 0.22777    \\ 
$c$         & 3.27473  & 2.70721  & 2.28554  & 1.85449  & 1.66757   & 1.52332     \\ \hline
\end{tabular}
\end{center}
\end{table}
%
%
The variational energies shown in Table~\ref{tabvar} 
are in excellent agreement with numerical diagonalization 
results reported in Table~\ref{tabdiag}.
This agreement holds for the whole range of 
Coulomb correlations and perpendicular magnetic fields 
considered in this work.

In the strong magnetic field limit, the variational energies
are practically identical (within the range of very small statistical errors)
to the exact numerical diagonalization values
indicating the overall excellent quality of the trial wave function.
Even more remarkable is the fact that for any combination of 
$\lambda$-s and $\gamma$-s the angular momentum of the 
lowest variational energy state is always reached at the exact value
obtained from the exact numerical diagonalizations.
We note there are cases, such as 
$\lambda=5$ and $\gamma=4$ where the energy difference between 
ground state and higher states with different angular 
momentum is extremely small.
( For $\lambda=5$ and $\gamma=4$ the diagonalization method
  gives a ground state with energy:
  $\epsilon=7.90109$ and angular momentum $|m_z|=4$,
  while the first excited state has an energy slightly higher:
  $\epsilon=7.90640$ and angular momentum $|m_z|=5$ ).
Nevertheless, in all occasions the trial wave function with lowest energy
has an angular momentum that corresponds to the exact diagonalization value.

Both diagonalization and variational results are in full 
agreement and confirm the expected outcome that:
(i) in absence of Coulomb correlations ($\lambda=0$) 
or 
(ii) in absence of a perpendicular magnetic field ($\gamma=0$),
the ground state has zero angular momentum.
However, when both $\lambda \neq 0$ and $\gamma \neq 0$, 
ground states with nonzero angular momentum arise.
%
For very large values of the perpendicular magnetic field the
ground state has increasingly large angular momentum values
where each change of $|m_z|$ indicates a singlet to triplet 
spin state transition, a phenomena that has been
observed in recent experiments~\cite{singlettriplet} 
with ultrasmall quantum dots.
Since the ground state spin of 2D quantum dot helium
can be either singlet ($S=0$) or 
triplet ($S=1$) it is plausible to expect that this quantum dot
has the potential to serve as qubit of a quantum computer, 
with the magnetic field tuning the transition between the 
two spin states, an idea
suggested by Burkard et al.~\cite{quantumgate}

Because the change of angular momentum indicates a
spin state transition, it is the mean square distance between the
two electrons (which is directly related to the angular momentum number)
that should indicate jumps or non-monotonic behavior contrary to the
dependence of energy on $\lambda$ that is monotonic.
Therefore, in addition to ground state energies, we also calculated the 
mean square distance between two electrons,
$\langle |\vec{\rho}_1-\vec{\rho}_2|^2 \rangle$
for a wide range of $\lambda$-s and $\gamma$-s.
The results are displayed in Table.~\ref{tabr2} where we show
$\alpha^2 \, \langle |\vec{\rho}_1-\vec{\rho}_2|^2 \rangle$ 
for values of 
$\lambda=0, 1, \ldots, 10 $ and 
$\gamma=0,1,\ldots 5$.
\begin{table}
\caption[!t]{ The optimal variational value of 
            $\alpha^2 \, \langle |\vec{\rho}_1-\vec{\rho}_2|^2 \rangle$
            for 2D quantum dot helium subject to a magnetic field
            as a function of dimensionless Coulomb coupling parameter, 
            $\lambda=e^2 \alpha/(4 \pi \epsilon_0 \epsilon_r \hbar \omega_0)=0,
             1, \ldots 10 $ and values of magnetic field,
            $\gamma=\omega_c/\omega_0=0,1,\ldots 5$.
            The parameter
            $ \alpha=\sqrt{m \, \omega_0/\hbar}$ has the dimensionality
            of an inverse length.   }
\label{tabr2}
\begin{center}
\begin{tabular}{|l|l|l|l|l|l|l|}
\hline                                                  
 & $\gamma=0$ & $\gamma=1$ & $\gamma=2$ & $\gamma=3$ 
           & $\gamma=4$ & $\gamma=5$ 
                                                        \\ \hline
$\lambda=0$ & 2.00000 & 1.78885 & 1.41421  & 1.10940  & 0.894427   
                                           & 0.742781 
                                                        \\ \hline
$\lambda=1$ & 3.18193 & 2.79350 & 3.19963  & 2.47708  & 1.97594  
                                           & 2.33444  
                                                        \\ \hline
$\lambda=2$ & 4.15428 & 4.62431 & 3.56731  & 3.71459  & 2.96353
                                           & 3.14825  
                                                        \\ \hline
$\lambda=3$ & 4.97151 & 5.12920 & 3.92383  & 3.90825  & 3.92870
                                           & 3.94601  
                                                        \\ \hline
$\lambda=4$ & 5.69236 & 5.61781 & 5.35004  & 5.08275  & 4.88226  
                                           & 4.73529  
                                                        \\ \hline
$\lambda=5$ & 6.34774 & 6.09048 & 5.62186  & 5.24316  & 4.98332
                                           & 5.51752  
                                                        \\ \hline
$\lambda=6$ & 6.95629 & 6.54753 & 5.89000  & 6.39368  & 5.91907
                                           & 5.58307  
                                                        \\ \hline
$\lambda=7$ & 7.52912 & 6.98994 & 7.27192  & 6.53384  & 6.85216
                                           & 6.35681  
                                                        \\ \hline
$\lambda=8$ & 8.07359 & 7.41855 & 7.49712  & 7.67346  & 6.93629
                                           & 7.12900  
                                                        \\ \hline
$\lambda=9$ & 8.59467 & 8.80371 & 7.72156  & 7.79918  & 7.86399
                                           & 7.18439  
                                                        \\ \hline
$\lambda=10$ & 9.09607 & 9.16053 & 9.08988  & 7.92518  & 7.94189  
                                            & 7.95454  
                                                       \\ \hline
\end{tabular}
\end{center}
\end{table}
\begin{figure}[!htb]
\includegraphics{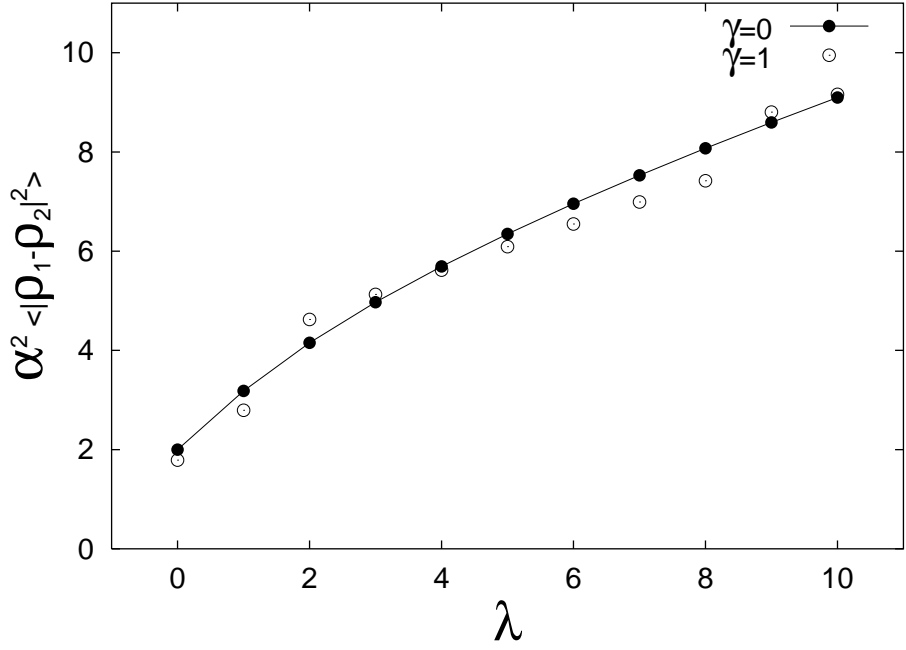}
\caption[]{ 
            Variational mean square distance between two electrons,
            $\alpha^2 \, \langle |\vec{\rho}_1-\vec{\rho}_2|^2 \rangle$
            for 2D quantum dot helium system in a
            perpendicular magnetic field
            as a function of dimensionless Coulomb coupling parameter, 
            $\lambda=e^2 \alpha/(4 \pi \epsilon_0 \epsilon_r \hbar \omega_0)$
            for two values of magnetic field corresponding to
            $\gamma=\omega_c/\omega_0=0$ and $1$.
            The line joining the data points at zero magnetic
            field serves as a guide to the eye.
            }
\label{figr2}
\end{figure}
In absence of Coulomb correlations,
the increase of the magnetic field brings electrons closer 
to each other resulting in a reduced mean square distance.
However, in presence of Coulomb correlations
there are values of $\lambda$ and $\gamma$ where electrons find
energetically favorable to jump to outer orbits ( increasing 
the angular momentum and their mean square distance, as well)
despite the effect of the magnetic field.
A manifestation of this behavior comes in the form of jumps of
            $\alpha^2 \, \langle |\vec{\rho}_1-\vec{\rho}_2|^2 \rangle$
for magnetic field $\gamma=1$ relative to the 
zero magnetic field case ($\gamma=0$) as seen in Fig.~\ref{figr2}.
For instance for Coulomb correlation values
$\lambda=0$ and $\lambda=1$ electrons are squized closer to
each other in presence of a magnetic field 
( the values of $\alpha^2 \, \langle |\vec{\rho}_1-\vec{\rho}_2|^2 \rangle$
  at $\gamma=1$ represented by empty circles are below the 
  corresponding values at $\gamma=0$ represented by filled circles), 
however for a larger correlation, such as $\lambda=2$ this is not the 
case any more.

The increase in the mean square distance between the two electrons reflects
the formation of a rotating electron(Wigner) molecule (REM or RWM).
Quantum mechanically one can immediately see that 
$\langle \vec{R} \rangle=0$, where $\vec{R}$ is the center of 
mass (CM) position
and
$\langle \vec{\rho}_1 \rangle=-\langle \vec{\rho}_2 \rangle$.
This represents the quantum counterpart of the classical minumum energy
configuration for two point charges in a harmonic trap that requires 
$\vec{\rho}_1=-\vec{\rho}_2$.
A plot of the electronic density will reveal that the electrons are mainly
localized on a thin ring centered at the zero of the parabolic potential.
The electronic density will exhibit the circular symmetry of the Hamiltonian 
and will only depend on distances (no angular dependence).
The electronic density is insensitive to angular correlations which are
very important, particularly for large values of the angular momentum.
A better probe of the angular characteristcs of the wave function is the 
conditional
probablity distribution (CPD) function~\cite{uzicpd} which for $N$ electrons 
is quite generally defined as:
$P(\vec{\rho},\vec{\rho}_0)=\langle \Psi | \sum_{i=1}^{N} \sum_{j \neq i}^{N}
\delta(\vec{\rho}_i-\vec{\rho}) \, \delta(\vec{\rho}_j-\vec{\rho}) | 
\Psi \rangle/\langle \Psi|\Psi \rangle$,
where $\Psi$ is the wave function under consideration.
For the case of quantum dot helium ($N=2$) this becomes:
\begin{equation}
P(\vec{\rho},\vec{\rho}_0)=2 \, \frac{ |\Psi(\vec{\rho},\vec{\rho}_0)|^2}{\langle \Psi|\Psi \rangle} \ ,
\label{cpd}
\end{equation}
where the wave function is given from Eq.(\ref{jastrow_laughlin}).
When calculating the CPD function the position vector of one 
electron, $\vec{\rho}_0$ is fixed, while $\vec{\rho}$
is moved so the resulting function of $\vec{\rho}$ measures the probability
of finding one electron at $\vec{\rho}$ given that there is one 
located at $\vec{\rho}_0$.
Obviously, for any choice of $\vec{\rho}_0 \neq 0$ the CPD function 
enables us to obtain the angular distribution of the second electron.
\begin{figure}[!tb]
\includegraphics{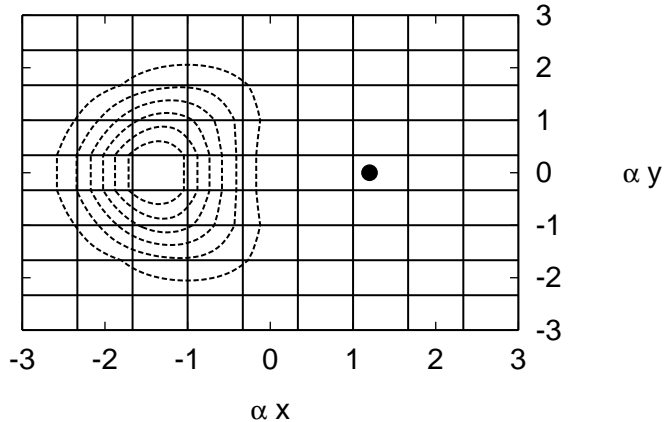}
\caption[]{
Contour plots of the CPD function, $P(\vec{\rho},\vec{\rho}_0)$
corresponding to the displaced Gaussian ground state wave function for
2D quantum dot helium at zero magnetic field ($\gamma=0$) and $\lambda=10$.
The black dot denotes the location of the fixed electron sittuated at position
$\alpha \, \vec{\rho}_0=\left( \alpha \, x_0 \neq 0, 0   \right)$. 
Distances are given in dimensionless units where $\alpha$ is the 
inverse oscillator length, $\sqrt{m \, \omega_0/\hbar}$.}
\label{figcpd}
\end{figure}
Fig.~\ref{figcpd} shows the CPD function for the ground state of
2D quantum dot helium at $\gamma=0$ and $\lambda=10$.
The contour plots of the CPD function shown in the present work are 
calculated with $\vec{\rho}_0$ on the $x$ axis with $\rho_0$
corresponding to the distance at which the electronic density
function (that is circularly ring shaped) has the maximum (crudely this
can be thought as the radius of the thin ring). 
The black dot indicates $\vec{\rho}_0$.
One can see clearly that the second electron is mainly localized
in the opposite position of the fixed electron.
This demonstrates that the quantum ground state has the same symmetry as the
classical lowest energy configuration.

\section{Discussion}
\label{discussion}

Beyond the ground-state energetics and ground-state angular momenta, we test
the accuracy of the trial wave function in two well known limits:
(i) infinite magnetic field limit ($\gamma \rightarrow \infty$) where the ground
    state energy should approach the energy of a classical system of $N=2$ point
    charges in a parabolic potential (adjusted by the quantum zero-point energy
    when Coulomb correlations are absent) and
(ii) lowest Landau level limit where the ground state energy of the trial wave function
     must coincide with the lowest Landau level Laughlin wave function for $N=2$
     electrons without the parabolic potential confinement.

To study limit (i) we calculate the lowest classical energy, $E_c$ for $N=2$ point charges
in a harmonic potential which in dimensionless units is:
\begin{equation}
\epsilon_{c}=\frac{E_c}{(\hbar \, \omega_0)}=\frac{3}{4} \, (2 \, \lambda)^{2/3} \ .
\label{eclassical}
\end{equation}
We note that the classical energy is not afected by the presence or absence of a magnetic field.
The classical ground state configuration for $N=2$ electrons is one in which the
respective positions of the particles are exactly opposite to each other at an
optimal distance, $\vec{\rho}_1=-\vec{\rho}_2 \neq 0$.

Naturally, one cannot immediately compare the quantum variational energy, $\epsilon$
to its classical counterpart, $\epsilon_c$ since without Coulomb correlations ($\lambda=0$)
the lowest quantum energy, $\epsilon$ is nonzero while the lowest classical energy, $\epsilon_c$
is zero.
Such difference in energy between the two quantities represents the quantum 
zero-point energy (at $\lambda=0$) which in
dimensionless units is:

\begin{equation}
\epsilon_0=\frac{E_0}{\hbar \, \omega_0}=2 \, \sqrt{1+\frac{\gamma^2}{4}} \ .
\label{ezero}
\end{equation}

\begin{figure}[!t]
\includegraphics{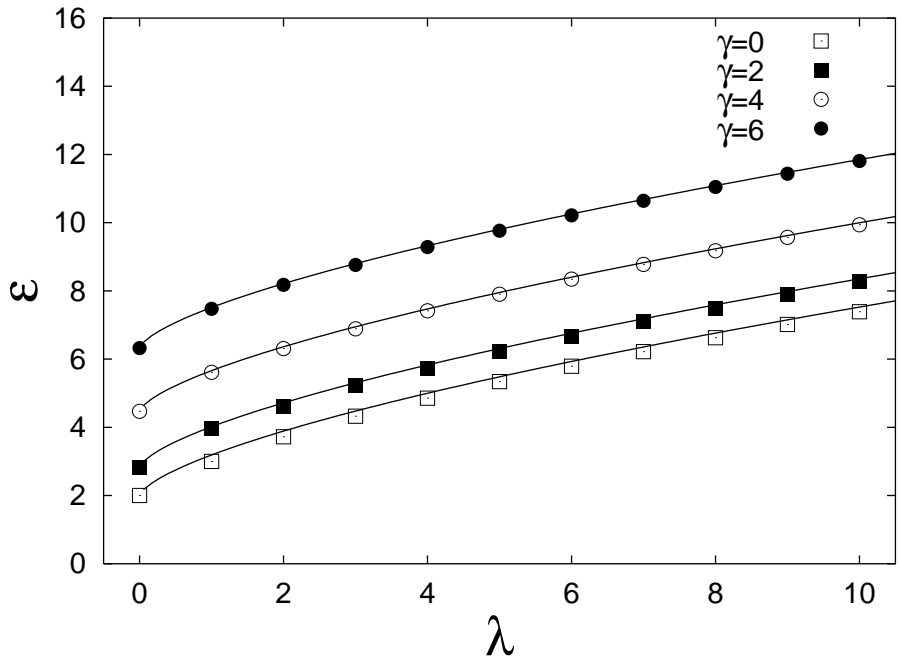}
\caption[]{ 
            Variational ground state energy
            of 2D quantum dot helium in a perpendicular magnetic field,
            $\epsilon=E/(\hbar \omega_0)$
            as a function of dimensionless Coulomb coupling parameter, 
            $\lambda=e^2 \alpha/(4 \pi \epsilon_0 \epsilon_r \hbar \omega_0)$
            for values of magnetic field corresponding to
            $\gamma=\omega_c/\omega_0=0, 2, 4,$ and $6$.
            The solid lines represents the "adjusted" classical energy,
            $\epsilon_c+\epsilon_0$ as a function of $\lambda$ calculated
            at the given $\gamma$ values.
            }
\label{figevar-lam}
\end{figure}

If we adjust the classical energy by $\epsilon_0$ the
quantities to compare are $\epsilon$ versus ($\epsilon_c+\epsilon_0$).
%
%
%
%
%
Fig.~\ref{figevar-lam} shows
the variational ground state energy of 2D quantum dot helium 
in a perpendicular magnetic field, $\epsilon$ and the adjusted
classical energy, $\epsilon_c+\epsilon_0$ (solid lines)
as a function of Coulomb correlation parameter, $\lambda$
for selected values of the magnetic field parameter, $\gamma$.
Quite surprisingly there is very good agreement between the "quantum" ground state energy
and the "adjusted classical" ground state energy at all magnetic fields including weak magnetic
fields.
As the magnetic field grows (increasing values of $\gamma$) the agreement only improves
as can clearly be seen from the data.
We also checked that in the infinite magnetic field limit
($\gamma \rightarrow \infty$) both $\epsilon(variational)-\epsilon_0$ and
$\epsilon(diag)-\epsilon_0$ tend to $\epsilon_c(classical)$.
This behavior was first noted by Yannouleas and Landman~\cite{uzi1}
in a study in which RWM wave functions were used
to describe few electron quantum dots.
They also recognized the importance of such finding in challenging the
composite fermion picture of quantum dots which instead implies that
$\epsilon-\epsilon_0 \rightarrow 0$ as $\gamma \rightarrow \infty$.
The same result is obtained here using not the RWM wave function,
but the displaced Gaussian variational wave function.

\begin{figure}[!t]
\includegraphics{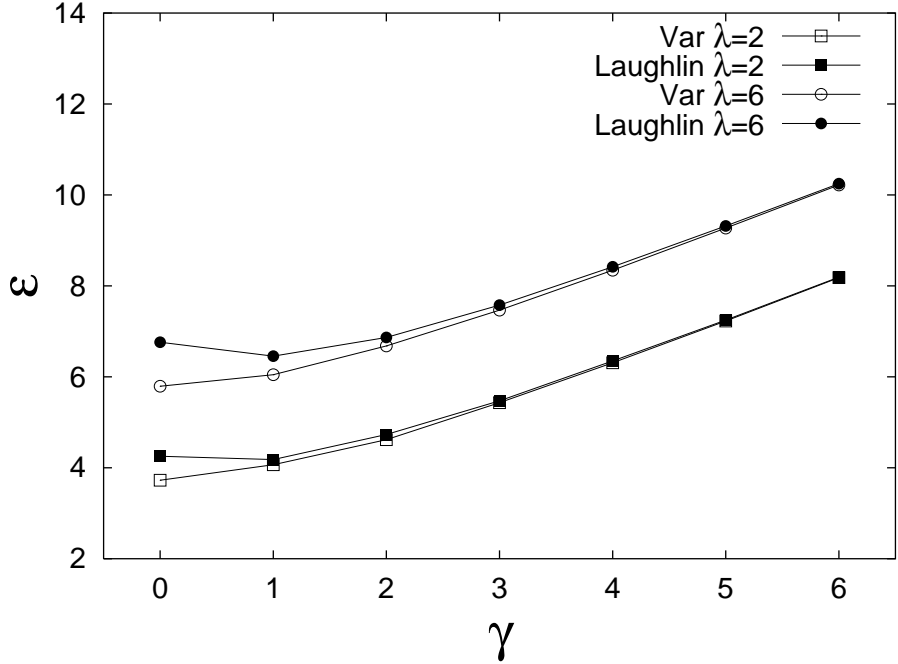}
\caption[]{ 
            The ground state energy curves, $\epsilon=E/(\hbar \omega_0)$ as a function
            of the magnetic field parameter, $\gamma$ for the case of
            displaced Gaussian variational wave function (Var) and
            Laughlin's wave function (Laughlin) for two values of the Coulomb
            correlation strength: $\lambda=2$ and $6$.
            The solid lines joining the data points serve as a guide to the eye.
            }
\label{figevar-elaughlin}
\end{figure}

To check limiting behaviour (ii) we use Laughlin wave function to calculate
the energies for the same $\lambda$-s and $\gamma$-s in which the displaced Gaussian
trial wave function was used and then compare the results.
The ground state energies obtained with the Laughlin
wave function are  displayed in
Table~\ref{tablaughlin}.
The angular momentum, $m_z$ for which the lowest
Laughlin energy is obtained is also specified.
\begin{table}[!t]
\caption[ ]{Ground state energies, $\epsilon=E/(\hbar \omega_0)$ 
            corresponding to the Laughlin wave function
            for 2D quantum dot helium subject to a perpendicular
            magnetic field for given values of dimensionless  
            parameters
            $\lambda$
             and
            $\gamma$
            The angular momentum, $m_z$ of the ground state is also
            specified. }
\label{tablaughlin}
\begin{center}
\begin{tabular}{|l|l|l|l|l|l|l|}
\hline                                                  

 & $\gamma=0$ & $\gamma=1$ & $\gamma=2$ & $\gamma=3$ 
           & $\gamma=4$ & $\gamma=5$                      \\ \hline
$\lambda=0$ & 2.00000 & 2.23607 & 2.82843  & 3.60555  & 4.47214   & 5.38516 
                                                            \\ 
  &$m_z=0$ & $m_z=0$ & $m_z=0$ & $m_z=0$ & $m_z=0$ & $m_z=0$  \\ \hline
$\lambda=1$  & 3.25331 & 3.51671 & 3.98787  & 4.74972  & 5.64527  & 6.54155 
                                                            \\ 
 &$m_z=0$ & $m_z=1$ & $m_z=1$ & $m_z=1$ & $m_z=1$ & $m_z=2$   \\ \hline
$\lambda=2$ & 4.25331 & 4.17932 & 4.73309  & 5.47320  & 6.34988  & 7.24827
                                                            \\ 
 & $m_z=1$ & $m_z=1$ & $m_z=1$ & $m_z=2$ & $m_z=2$ & $m_z=3$ \\ \hline
$\lambda=3$ & 4.87997 & 4.84193 & 5.33361  & 6.09150  & 6.93735  & 7.84253 
                                                            \\ 
            & $m_z=1$ & $m_z=1$ & $m_z=2$ & $m_z=3$ & $m_z=3$ & $m_z=4$  \\ \hline
$\lambda=4$ & 5.50663 & 5.45996 & 5.89253  & 6.61737  & 7.46625  & 8.37252 
                                                            \\ 
            & $m_z=1$ & $m_z=2$ & $m_z=2$ & $m_z=3$ & $m_z=4$    & $m_z=5$   \\ \hline
$\lambda=5$ & 6.13329 & 5.95692 & 6.39990  & 7.11735  & 7.95855  & 8.86033 
                                                            \\ 
            & $m_z=1$ & $m_z=2$ & $m_z=3$ & $m_z=4$ & $m_z=5$ & $m_z=6$  \\ \hline
$\lambda=6$ & 6.75994 & 6.45388 & 6.86566  & 7.57749  & 8.41976  & 9.31802 
                                                            \\ 
            & $m_z=1$ & $m_z=2$ & $m_z=3$ & $m_z=4$ & $m_z=5$ & $m_z=7$  \\ \hline
\end{tabular}
\end{center}
\end{table}
As expected, one notes that Laughlin wave function is not a good description
of the system at weak (and zero) magnetic fields (for instance, 
there are several occasions
in which the Laughlin ground state energy has the wrong angular
momentum, such as the cases $\gamma=0$ and $\lambda=2, 3, \ldots $ etc.).
However, comparing the energies in Table~\ref{tabvar} to Table~\ref{tablaughlin} 
one sees that, in the limit of strong magnetic fields (increasing $\gamma$-s),
the ground state energy of the displaced Gaussian variational wave function quickly
approaches the Laughlin values.
This is clearly seen in Fig.~\ref{figevar-elaughlin} where we plot the ground state energy 
corresponding to the displaced Gaussian variational wave function and the
Laughlin wave function for $\lambda=2$ and $6$ as a function of
the magnetic field parameter, $\gamma$. 
The convergence of variational and Laughlin ground state energies in the limit
of strong magnetic field is quite general and happens at any arbitrary value of
the Coulomb correlation strength, $\lambda$, though for clarity of the plot 
in Fig.~\ref{figevar-elaughlin} we display only the curves corresponding to $\lambda=2$ and $6$.

Given the very good performance of the trial wave function on
describing 2D quantum dot helium in an arbitrary perpendicular
magnetic field, we plan to generalize the treatment to larger quantum dots.
In this case the liquid or solid (crystalline) character of larger quantum dots 
in a magnetic field is rather non trivial and crucially depends on the values
of $\gamma$, $\lambda$ as well as density of the system.
To that effect we would describe any 2D quantum dot in a perpendicular 
magnetic field with a generalized trial wave function of the form: 
\begin{equation}
\Psi_{N}=
\prod_{i<j}^N \left[ J(\rho_{ij}) \times (z_i-z_j)^{n} \right] \, 
D_{\uparrow}(\Phi) D_{\downarrow}(\Phi) \,
 \chi(S)  \ , 
\label{generalwf}
\end{equation}
where $N$ is the number of electrons in the dot,
$ \chi(S)=\chi(s_1,s_2,\ldots s_N)$ is the spin function
for $N_{\uparrow}$ spin-up and 
$N_{\downarrow}=N-N_{\uparrow}$ spin-down electrons
and 
the space wave function has a Jastrow-Slater form.
The determinants
$D_{\uparrow}(\Phi)$
and
$D_{\downarrow}(\Phi)$ 
are Slater determinants for spin-up
and spin-down electrons built out of:
(i) FD orbitals~\cite{fock,darwin} or
(ii) Gaussian localized orbitals~\cite{nosanow} (for a crystalline state, only).
The displaced Gaussian pair correlation factor,
$J(\rho_{ij})$ as specified in Eq.(\ref{jasmanybodywf}) gurantees 
the quality of the wave function at all magnetic fields ranging from
weak (and zero) to strong 
and
the integer quantum number, $n=0,1,2,\ldots$
which takes even/odd values for respectively 
antisymmetric/symmetric spin functions 
determines the overall parity of the space wave function
as required by the Pauli's principle.
In such a case, a full VMC simulation would be the method of choice
and the optimized trial wave function
can be further  used as a guiding function for DMC calculations.

\section{Conclusions}
\label{summary}

To conclude, we have introduced a very accurate trial wave function 
for 2D quantum dot helium in an arbitrary perpendicular magnetic field.
A key novel element of this description is a Jastrow pair correlation factor
that has a displaced-Gaussian form and contains two variational
parameters to optimize.
The variational energies are in excellent agreement with
exact numerical diagonalization calculations at any value of the
perpendicular magnetic field including weak (and zero) or strong fields.
In agreement with the RWM formalism we find that, for given values of the
Coulomb correlation strength, the quantum ground state energy 
of 2D quantum dot helium at any value of the magnetic field is close to the
value of the "adjusted" classical energy.
Quantum and classical energies converge in the limit of infinite magnetic field.
For a given Coulomb correlation strength and in the limit of infinite
magnetic field, the energies of the displaced Gaussian trial wave function 
agree very well with the energies obtained from Laughlin's wave function, 
though we note that Laughlin's wave function is a poor description of the system
for weak (zero) and intermediate magnetic fields.
For weak (zero) and intermediate magnetic fields a Jastrow pair correlation factor of the
nature studied in this work should be included in the total wave function
in conjuction with the Laughlin or RWM component, which 
are most effective in high magnetic fields.
A Jastrow pair correlation factor, such as the displaced Gaussian factor introduced 
in this study, is essential to provide an accurate description of the system at 
any value of the magnetic field not limited to high magnetic fields only.
Other Jastrow pair correlation factors such as those constructed through
Pade approximations with several variational parameters~\cite{pederiva} also
provide quite an accurate description of the system and compare favorably with
exact diagonalization results.
However, the displaced Gaussian pair correlation factor is the most
intuitive and the simplest physical choice that guarantees a consistent and excellent description 
of quantum dot helium system at all magnetic fields ranging from weak (zero) to infinity.
A generalization of this trial wave function for $N$-electron 
quantum dots in a perpendicular magnetic field is also discussed.
%


\section{Acknowledgments}

The authors thank A. Anil Kumar for useful discussions.
This research was supported by the U.S. D.O.E. (Grant No. DE-FG52-05NA27036)
and 
by the Office of the Vice-President
for Research and Development of Prairie View A\&M University through
a 2003-2004 Research Enhancement Program grant.


\end{document}